\documentclass[10pt,journal,compsoc]{IEEEtran}

\usepackage{cite}
\usepackage{amsmath,amssymb,amsfonts}
\usepackage{algorithmic}
\usepackage{graphicx}
\usepackage{textcomp}
\usepackage{xcolor}

\usepackage{multirow}
\usepackage[utf8]{inputenc}
\usepackage{amsmath,amssymb,amsfonts}
\usepackage{algorithmic}
\usepackage{algorithm}
\usepackage{graphicx}
\usepackage{textcomp}
\usepackage{color,xcolor}
\usepackage{url}
\usepackage{graphicx}
\usepackage{subfigure}
\usepackage{stmaryrd}
\usepackage{caption}
\usepackage{verbatimbox}
\usepackage{enumerate}

\usepackage{makecell}
\usepackage{booktabs}
\newcommand{\tabitem}{\textbullet~~}
\usepackage{multirow}
\AtBeginDocument{%
  \providecommand\BibTeX{{%
    \normalfont B\kern-0.5em{\scshape i\kern-0.25em b}\kern-0.8em\TeX}}}
\usepackage{enumitem}
\usepackage{colortbl}
\definecolor{codegreen}{rgb}{0,0.6,0}
\definecolor{codegray}{rgb}{0.5,0.5,0.5}
\definecolor{codepurple}{rgb}{0.58,0,0.82}
\definecolor{backcolour}{rgb}{0.95,0.95,0.92}
\usepackage{listings}
\lstdefinestyle{mystyle}{
  backgroundcolor=\color{backcolour},   commentstyle=\color{codegreen},
  keywordstyle=\color{magenta},
  numberstyle=\tiny\color{codegray},
  stringstyle=\color{codepurple},
  basicstyle=\ttfamily\footnotesize,
  breakatwhitespace=false,
  breaklines=true,
  captionpos=b,
  keepspaces=true,
  numbers=left,
  numbersep=5pt,
  showspaces=false,
  showstringspaces=false,
  showtabs=false,
  tabsize=2
}
\newcommand{\finding}[2]{
\begin{center}
\fcolorbox{black}{gray!10}{\parbox{.97\linewidth}{
\textbf{Answer to RQ{#1}:}
{#2}
}}
\end{center}
}

\begin{document}

\title{Program Repair: Automated vs. Manual}

\author{Quanjun Zhang, Yuan Zhao, Weisong Sun, Chunrong Fang*, Ziyuan Wang*, Lingming Zhang

\IEEEcompsocitemizethanks{
\IEEEcompsocthanksitem 
Quanjun Zhang, Yuan Zhao Weisong, Sun and Chunrong Fang 
are with the State Key Laboratory for Novel Software Technology, Nanjing University, China. \protect\\
E-mail: 
quanjun.zhang@smail.nju.edu.cn,
allenzcrazy@gmail.com,
weisongsun@smail.nju.edu.cn,
fangchunrong@nju.edu.cn

\IEEEcompsocthanksitem Ziyuan Wang is with Nanjing University of Posts and Telecommunications, China. \protect\\
E-mail: wangziyuan@njupt.edu.cn

\IEEEcompsocthanksitem Lingming Zhang is with Department of Computer Science, University of Illinois at Urbana-Champaign, USA. \protect\\
E-mail: lingming@illinois.edu

\IEEEcompsocthanksitem *Chunrong Fang and Ziyuan Wang are corresponding authors.

}
}

\markboth{Journal of \LaTeX\ Class Files,~Vol.~14, No.~8, August~2015}%
{Shell \MakeLowercase{\textit{et al.}}: Bare Demo of IEEEtran.cls for Computer Society Journals}



\IEEEtitleabstractindextext{
\begin{abstract}

Various automated program repair (APR) techniques have been proposed to fix bugs automatically in the last decade.
Although recent researches have made significant progress on the effectiveness and efficiency,
it is still unclear how APR techniques perform with human intervention in a real debugging scenario.
To bridge this gap, we conduct an extensive study to compare three state-of-the-art APR tools with manual program repair, and further investigate whether the assistance of APR tools (i.e., repair reports) can improve manual program repair.
To that end, we recruit 20 participants for a controlled experiment, resulting in a total of 160 manual repair tasks and a questionnaire survey.
The experiment reveals several notable observations that
(1) manual program repair may be influenced by the frequency of repair actions sometimes;
(2) APR tools are more efficient in terms of debugging time, while manual program repair tends to generate a correct patch with fewer attempts;
(3) APR tools can further improve manual program repair regarding the number of correctly-fixed bugs, while there exists a negative impact on the patch correctness;
(4) participants are used to consuming more time to identify incorrect patches, while they are still misguided easily;
(5)
participants are positive about the tools' repair performance, while they generally lack confidence about the usability in practice.
Besides, we provide some guidelines for improving the usability of APR tools (e.g., the misleading information in reports and the observation of feedback).


\end{abstract}

\begin{IEEEkeywords}
automated debugging, automated program repair, human study, manual program repair
\end{IEEEkeywords}
}

\maketitle
\IEEEdisplaynontitleabstractindextext

\IEEEpeerreviewmaketitle

\IEEEraisesectionheading{
\section{Introduction}
\label{sec:introduction}
}
\IEEEPARstart{M}odern software systems continuously evolve with prevalent bugs, which have been widely recognized as notoriously costly and disastrous \cite{2019Gazzola}.
Manual debugging can be an extremely time-consuming and error-prone task due to the increasing size and complexity of software systems \cite{2002Planning, 2005Anvik, 2009Aranda}.
 For example, shown in a prior report, software debugging often accounts for over 50\% of the development cost of a software product \cite{2016Software}, and consumes billions of dollars globally every year \cite{2013Boulder}.
Therefore, a vast body of research effort has been dedicated to automated debugging, such as automated fault localization \cite{2019Lou} and automated program repair (APR) \cite{2019Ghanbari, 2018Xiong}.
The former aims to directly localize software buggy elements to alleviate manual effort, while the latter aims to automatically fix software bugs without human intervention.


Despite an emerging research area, APR has been extensively studied in the literature and recent researches have made significant progress on effectiveness and efficiency \cite{2018Monperrus}.
According to a living review of APR research \cite{2018MonperrusLiving}, multiple papers get published each year, introducing various delicately implemented APR tools.
Among them, numerous studies evaluate APR tools on the effectiveness, in terms of the number of correctly-fixed bugs \cite{2017Xiong, Jiang2018, 2019Ghanbari, 2019Tufano}.
For example, Ghanbari et al. \cite{2019Ghanbari} propose a novel tool PraPR based on mutation testing, which can successfully fix 43 real bugs from Defects4J \cite{2014Just}.
Other advanced tools, such as ACS \cite{2017Xiong} and SimFix \cite{Jiang2018}, have also produced promising results in terms of the number of bugs that can be fixed.
Another performance aspect that deserves investigation is the efficiency, in terms of repair time  and the number of patch candidates \cite{2019Ghanbari,  2020Liu}.
Recently, Kui et al. \cite{2020Liu} calculate the number of generated patch candidates before fixing a given bug to assess the efficiency.

Existing studies usually focus on the performance of the repair approaches regarding some criteria (e.g., repairability, correctness, and repair time) \cite{2019Durieux, 2020Liu}.
APR is indeed a growing field, and it is essential to understand how it can be applied in practice when developers get involved.
A natural question thus arises from the above scenario: \emph{How do state-of-the-art APR tools perform comparing with manual program repair}?
However, although APR has been extensively studied and even has drawn attention from industry (e.g., Facebook \cite{2019Bader, 2019Marginean} and Google \cite{2019Mesbah}), direct deployment of APR tools in the industry seems to consume constant research effort.
For example, research has identified that generated patches might be less readable and maintainable, even meaningless sometimes \cite{2012Fry, 2012Ceccato}.
Meanwhile, it is fundamentally difficult to achieve high precision for generated patches due to the weak test suites \cite{2018Monperrus,2015Qi,2018Yi}.
Besides, as repair reports generated by APR tools may provide suspicious elements and corresponding candidate fixes, developers could reduce repair effort with the assistance of these reports.
Rather than direct deployment, such a semi-automatic approach may be a more feasible application of APR tools at this point.
As such, \emph{are state-of-the-art APR tools beneficial to manual program repair as debugging aids}?

In this paper, we perform a large-scale human study to bridge the current gap.
Specifically, 8 real bugs with varied symptoms from the widely studied Defects4J benchmark and 3 state-of-the-art APR tools from all possible repair categories are randomly selected based on some well-designed criteria.
As far as we are aware, this is the largest user study in APR field ever (e.g, there exist 6 bugs and 2 APR tools in \cite{2014Tao}).
We also recruit 20 participants to individually fix all the bugs without or with different debugging aids (i.e., 3*8 repair reports generated by APR tools), resulting in a total of 160 manual program repair tasks.
Besides, a platform, namely MoocTest, with user-friendly debugging assistance, is also implemented, which can assign the bugs and corresponding debugging aids to every participant automatically.


Our work reveals several notable observations.
Firstly, manual program repair may be influenced by the frequency of repair actions (i.e., addition, removal and modification over code elements), while APR tools are effective for the bugs with certain types due to their design mechanisms.
Moreover, patches generated by manual program repair usually have a higher correctness ratio and code quality than those generated by APR tools.
Meanwhile, for successful repair tasks, APR tools are more efficient in terms of repair time, while participants need fewer attempts to generate a correct patch.

Secondly, we analyze whether manual program repair benefits from APR tools.
Our analysis confirms that APR tools indeed improve manual program repair regarding the number of correctly-fixed bugs, while the tools have a negative impact on the patch correctness and may not help reduce debugging time.
Besides, the types of repair tools (i.e., heuristic-based, constraint-based and template-based) have little influence on the patch correctness and debugging time, while the types of repair reports (log report, correct report and incorrect report) have a more significant impact on the patch correctness.
Further, when provided with incorrect patches, participants tend to consuming more debugging time, while they are still easily misguided.


Finally, we qualitatively investigate participants' opinions on the assistance of APR tools.
There is a huge gap between the repair performance in the benchmark and the usability in practice for state-of-the-art APR tools.
Besides, the feedback in log output should be valued more in the future.
Moreover, participants are positive about the quick identification of buggy elements and candidate fixes,
while they generally lack confidence in the accuracy and format of repair reports.

Based on these observations, we provide various practical guidelines on how to improve the usability of APR tools so as to benefit manual program repair further.
For example, we discuss how to present repair reports in a more user-friendly way and reduce misleading repair information.
Above all, we suggest that APR tools should pay more attention to the feedback in the log output, which can provide useful guidelines for all possible bugs.
Overall, our contributions are summarized as follows.
\begin{itemize}

      \item We conduct the first systematic large-scale human study to compare manual program repair against three state-of-the-art APR tools, and analyze whether the assistance of APR tools can improve manual program repair.
      \item We perform a questionnaire survey to analyze participants’ attitudes toward the assistance of APR tools, and further provide various practical guidelines on how to improve the usability of APR tools in practice.
      \item We release all experimental data (including all raw data, experimental script, and result analysis) for replication and future research on APR at {https://github.com/AnonymousAPR/Data}.
\end{itemize}

\section{Related Work and motivation}
\subsection{Automated program Repair}
\subsubsection{Typical repair steps}
\label{apr_types}
 The primary objective of APR techniques is to identify and fix the bug automatically.
A typical repair technique is usually composed of three steps:
(1) off-the-shelf fault localization techniques are used to recognize the suspicious code elements \cite{2007Abreu,  2008Ayewah,2016Wong};
(2) these elements are then modified based on a set of transformation rules to generate various new program variants (i.e., candidate patches);
(3) the original test suite is adopted as the oracle to verify all candidate patches.
Specifically, a candidate patch passing the original test suite is called a \emph{plausible} patch.
A plausible patch, which is also semantically equivalent to the developer patch, denotes a \emph{correct} patch.

\begin{figure}[htbp]
\centering
\graphicspath{{graphs/}}
    \includegraphics[width=0.47\textwidth]{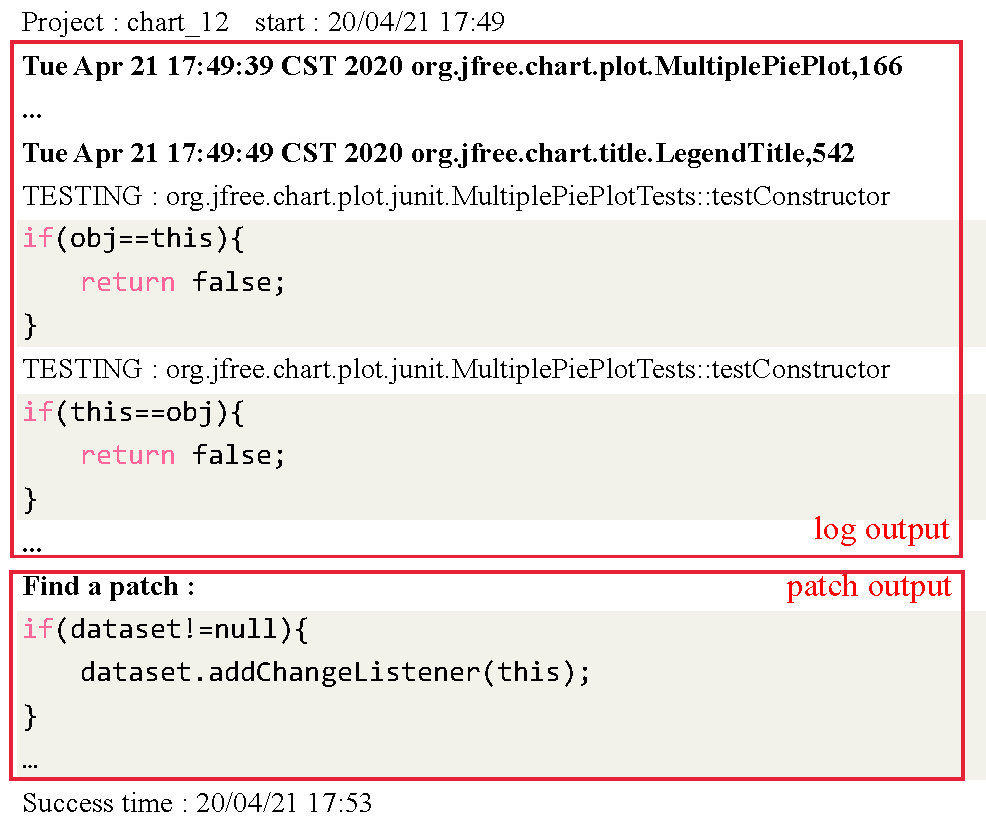}
    \caption{A typical repair report generated by SimFix}
    \label{fig:report}
\end{figure}

In practice, a typical repair report generated by an APR tool can be found in Fig. \ref{fig:report}.
Specifically, the report is usually divided into two parts (i.e., the log output and patch output), where the former contains the log information returned by the first two steps (i.e., the suspicious statements and candidate patches already tried), and the latter contains the patch information returned by the last step (i.e., the plausible patch passing all the available test cases).
As a reminder, the repair report will only contain the log output if the APR tool cannot generate a plausible patch for a given bug.

\subsubsection{Patch generation technique}
In the literature, a variety of techniques \cite{2016Ji, 2012Logozzo, 2015long, 2016long, 2014Pei, 2015Tan, 2018Wen, 2017Xuan} are adopted to generate patches based on different mechanisms.
Following recent APR work \cite{2020Liu,2020Benton,2020Wang}, we categorize them into four main categories:
heuristic-based, constraint-based, template-based and learning-based repair techniques.

\emph{$\bullet$ Heuristic-based repair techniques.}
These techniques usually use a heuristic algorithm to find a valid patch by iteratively exploring a search space of syntactic program modifications \cite{Le2011, 2016Martinez, 2018Yuan}.
Among them, GenProg \cite{Le2011} proposed in the early days has been considered a seminal work in this field, which performs delete and replace operations at the abstract syntax tree (AST) level.
The recent SimFix technique \cite{Jiang2018} utilizes code change operations from existing patches and similar code snippets to build two search spaces.
Then, the intersection of the above two search spaces is further used to search the final patch by basic heuristics.

\emph{$\bullet$ Constraint-based repair techniques.}
These techniques mainly focus on conditional statement repair, which can repair more than half of the bugs in existing approaches \cite{2018Martinez, 2016Durieux, 2016Xuan}.
In detail, these techniques transform the patch generation into a constraint solving problem, and use a solver to obtain a feasible solution.
For example, Nopol \cite{2016Xuan} relies on an SMT solver to solve the condition synthesis problem after identifying potential locations of patches by angelic fix localization and collecting test execution traces of the program.
Among them, ACS \cite{2017Xiong} proposed to refine the ranking of ingredients for condition synthesis is considered as one of the most advanced constraint-based repair techniques \cite{2020Liu}.

\emph{$\bullet$ Template-based repair techniques.}
These techniques generate patches by mutating a buggy program, similar to mutation testing \cite{2018Zhang}.
In detail, these techniques generate patches by following fix patterns to mutate buggy code entities with retrieved donor code \cite{2020Koyuncu, 2019Liu, 2019LiuAvatar}.
For example, Liu et al. \cite{2019Liu} revisit the repair performance of fix patterns via a systematic study assessing the effectiveness of a variety of fix patterns summarized from the literature.
Among them, the most recent technique PraPR \cite{2019Ghanbari} is able to generate plausible and correct patches for 148 and 43 real bugs, respectively, which is the largest number of bugs reported as fixed for Defects4J when published.

\emph{$\bullet$ Learning-based repair techniques.}
These techniques attempt to fix bugs enhanced by machine learning techniques \cite{2019White, 2019Tufano, 2017Gupta, 2020Lutellier, 2020Li, 2021Jiang}.
For example, Tufano et al. \cite{2019Tufano} extensively evaluate the ability of adopting neural machine translation (NMT) techniques to generate patches from bug-fixes commits in the wild.
Furthermore, Lutellier et al. \cite{2020Lutellier} propose a new context-aware NMT architecture that represents the buggy source code and its surrounding context separately, to automatically fix bugs in multiple programming languages.
Unlike techniques in above three categories, learning-based techniques generally require extra training data (i.e, the tuples of buggy, context, and fixed lines of code) to capture complex relations between buggy and fixed code.

In our work, we initially aim to consider all existing APR tools.
However, due to the fact that our human study involving too many tools may be unaffordable,
we select several representative APR tools from the above categories based on the criteria (Section \ref{apr_selection}) for our evaluation.

%
%

\subsection{Automated debugging with human intervention}
\label{related work with human study}
Various automated debugging techniques have been proposed over the past two decades \cite{2008Ayewah,2016Wong}.
Furthermore, various empirical studies involving human intervention are also conducted to evaluate the usefulness of automated debugging techniques \cite{2011Parnin, 1986Weiser, 2015Wang}.

In similar areas of fault localization,
some studies have been conducted to analyze the performance of program debugging with or without human intervention \cite{2011Parnin, 2016Xie, 1986Weiser, 2001Francel, 2015Wang}.
Among them, SBFL utilizes testing outcomes to evaluate the suspiciousness for each program unit, and then they are ranked in descending order and provided to programmers as the suggested fault location.
Besides, Parnin et al. \cite{2011Parnin} require programmers (i.e., students) to perform debugging tasks with or without suspicious statements given by spectrum-based fault localization (SBFL) techniques.
By addressing some significant problems that are not resolved in the above study, Xie et al. \cite{2016Xie}
focus on the reasons behind the observed results via a quantitative focus-tracking analysis
to revisit the actual helpfulness of SBFL.


In the field of APR, Tao et al. \cite{2014Tao} investigate the usefulness of automatically generated patches as debugging aids,
where ten plausible patches (i.e., passing all the corresponding test cases) are selected from GenProg \cite{Le2011} and PAR \cite{2013Kim}.
However, they focus on the impact of selected patches' quality and do not concern other feedback (e.g., the fixes already tried and patch execution information) in the repair reports.
To date, even state-of-the-art APR tools can only fix a small ratio of real bugs (i.e., $<20\%$ for Defects4J \cite{2019Lou, 2017Xiong, 2019Ghanbari}), and such tools seem useless for the vast majority of unfixed bugs.
Recently, Lou et al. \cite{2019Lou} confirm that the feedback (i.e., the patch execution information) can provide useful guidelines for powerful fault localization, even when the APR tool fails to return a plausible patch for a bug.

\begin{table}[t]
\scriptsize
\centering
 \caption{Five top-ranked suspicious statements for Time-15 }
  \label{rankeds-statements}
  \ttfamily
  \setlength\tabcolsep{3pt}
    \begin{tabular}{c|c|c}
     \hline
        \textbf{EID} & \textbf{Statements Signature} & \textbf{\#Patch} \\ \hline
        $s_1$ & org.joda.time.field.FieldUtils,140 & 0 \\ \hline
        \rowcolor[rgb]{0.75,0.75,0.75}
        $s_2$ & org.joda.time.field.FieldUtils,138 & 1 \\ \hline
        $s_3$ & org.joda.time.field.FieldUtils,142 & 1 \\ \hline
        $s_4$ & org.joda.time.field.FieldUtils,144 & 81 \\ \hline
        $s_5$ & org.joda.time.field.FieldUtils,145 & 0 \\
     \hline
  \end{tabular}
\end{table}

Our insight is that such repair feedback can also provide useful debugging hints for repair tasks, which can be adopted by developers directly.
For example, when we apply state-of-the-art APR technique, SimFix, on a real bug Time-15 (i.e., denotes the 15th buggy version of Joda-Time project \cite{JodaTime}
from Defects4J, no plausible patch is produced.
Table \ref{rankeds-statements} lists the top-5 most suspicious statements and the number of patch candidates already tried.
SimFix generates only one candidate patch for the buggy statement $s_2$, while more than 80 attempts are made for the statement $s_4$.
Although no plausible patch is returned, we still observe other useful feedback for repairing the bug in the repair report.
For example, there exist 13 candidate patches for $s_4$ containing the same code snippet in Listing \ref{code}, which is very similar to the correct patch in Table \ref{Table1}.
Thus instead of focusing on the limited selected patches, our study considering the whole repair report, extends the evaluation scope of APR to all possible bugs.


\begin{lstlisting}[language=Java, caption=A code snippet in the log output, label=code]
if(total/val1!=val1||val1==Long.MIN_VALUE&&val2==-1||val2==Long.MIN_VALUE&&val1shi==-1){
    throw new ArithmeticException("Multiplication overflows a long: "+val1);
}
\end{lstlisting}

Besides, different APR tools vary dramatically in settings (i.e., the patch generation strategy and report format), which may affect programmers’ repair performance.
For example, ACS and SimFix return only one most plausible patch in source-code level while PraPR returns all plausible patches in bytecode level for each bug.
Thus multiple state-of-the-art APR tools from different categories are selected in our evaluation.
Based on the above insights, we present an extensive study to compare the performance of manual program repair with state-of-the-art APR tools and investigate whether the assistance of such tools benefits manual program repair.

\begin{table*}[htbp]
 \scriptsize
\centering
 \caption{The details of selected bugs}

  \label{Table1}
    \begin{tabular}{wm{1cm}|@{\hskip3pt}p{10.8cm}|p{5cm}|m{0.5cm}|m{0.5cm}|}
    \hline
    BugID & Buggy code snippet and Developer patch & Symptom and illustration \\
    \hline
    \hline

    Time-7 &
    \ttfamily
    \mdseries
    \setlength\tabcolsep{2pt}
    \begin{tabular}{lll}
        1 & & long instantMillis = instant.getMillis(); \\
        2 & & Chronology chrono = instant.getChronology(); \\
        3 & \textcolor{green}{+} & int defaultYear = DateTimeUtils.getChronology(chrono).year() \\
          & & .get(instantMillis); \\
        4 & & long instantLocal = instantMillis + chrono.getZone() \\
        4 & & .getOffset(instantMillis); \\
        5 & & chrono = selectChronology(chrono); \\
        6 & \textcolor{red}{-} & int defaultYear = chrono.year().get(instantLocal); \\
    \end{tabular}
    &
    \begin{minipage} [htbp]{5cm}
      \tabitem IllegalFieldValueException is thrown as wrong year could be obtained in non UTC zones. \\
      \tabitem
      The developer patch obtains correct year using the chronology of the ReadWritableInstant.
    \end{minipage}

    \\
    \hline

    Time-15 &
    \setlength\tabcolsep{2pt}
    \ttfamily

    \begin{tabular}{lll}
        1 & & switch (val2) \{ \\
        2 & & case -1: \\
        3 & \textcolor{green}{+} & if (val1 == Long.MIN\_VALUE) \{ \\
        4 & \textcolor{green}{+} & \qquad throw new ArithmeticException("Multiplication overflows a long: " \\
        && + val1 + \qquad " * " + val2); \\
        5 & \textcolor{green}{+} & \} \\
        6 & & return -val1; \\
    \end{tabular}
    &
    \begin{minipage} [htbp]{5cm}
        \tabitem The buggy code snippet cannot detect the overflow if the long val1 == Long.MIN\_VALUE and the int scalar == -1. \\
        \tabitem The developer patch fixes the bug by throwing an exception when the values of above two variables are incorrect.
    \end{minipage}
    \\
    \hline

    Lang-7 &
    \ttfamily
    \setlength\tabcolsep{2pt}
    \begin{tabular}{lll}
        1 & \textcolor{red}{-} & if (str.startsWith("--")) \{ \\
        2 & \textcolor{red}{-} & \qquad return null; \\
        3 & \textcolor{red}{-} & \} \\
        4 & & ... \\
        5 & \textcolor{green}{+} & if (str.trim().startsWith("--")) \{ \\
        6 & \textcolor{green}{+} & \qquad throw new NumberFormatException(str + " is not a valid number."); \\
        7 & \textcolor{green}{+} & \} \\
    \end{tabular}
    &
    \begin{minipage} [htbp]{5cm}
        \tabitem The buggy method createBigDecimal() would return null when the string starting with "--" is given, which is contrary to the behaviour of other methods and would throw NumberFormatException. \\
        \tabitem The developer patch fixes the bug by throwing an appropriate exception in the method createNumber() when the above string is given.
    \end{minipage}
    \\
    \hline

    Lang-10 &
    \ttfamily
    \setlength\tabcolsep{2pt}
    \begin{tabular}{lll}
        1 & \textcolor{red}{-} & boolean wasWhite= false; \\
         &  & ...\\
        2 & \textcolor{red}{-} & if(Character.isWhitespace(c)) \{ \\
        3 & \textcolor{red}{-} & \qquad if(!wasWhite) \{ \\
        4 & \textcolor{red}{-} & \qquad \qquad wasWhite= true; \\
        5 & \textcolor{red}{-} & \qquad \qquad regex.append("$\backslash\backslash$s*+"); \\
        6 & \textcolor{red}{-} & \qquad \} \\
        7 & \textcolor{red}{-} & \qquad continue; \\
        8 & \textcolor{red}{-} & \} \\
        9 & \textcolor{red}{-} & wasWhite= false; \\
    \end{tabular}
     &
    \begin{minipage} [htbp]{5cm}
        \tabitem The buggy code treats white-space specially, which is in conflict with other methods in the program. \\
        \tabitem The developer patch fixes the bug by commenting the relevant code.
    \end{minipage}
    \\
    \hline

    Math-5 &
    \ttfamily
    \setlength\tabcolsep{2pt}
    \begin{tabular}{lll}
        1 & & if (real == 0.0 \&\& imaginary == 0.0) \{ \\
        2 & \textcolor{red}{-} & \qquad return NaN; \\
        3 & \textcolor{green}{+} & \qquad return INF; \\
        4 & & \} \\
    \end{tabular}
     &
    \begin{minipage} [htbp]{5cm}
        \tabitem The buggy code returns the wrong value, and throws AssertionFailedError in line 2. \\
        \tabitem The developer patch fixes the bug by returning correct value.
    \end{minipage}
    \\
    \hline

    Math-73 &
    \ttfamily
    \setlength\tabcolsep{2pt}
    \begin{tabular}{lll}
        1 & \textcolor{green}{+} & if (yMin * yMax > 0) \{ \\
        2 & \textcolor{green}{+} & \qquad throw MathRuntimeException.createIllegalArgumentException( \\
        3 & \textcolor{green}{+} & \qquad NON\_BRACKETING\_MESSAGE, min, max, yMin, yMax); \\
        4 & \textcolor{green}{+} & \} \\
    \end{tabular}
    &
    \begin{minipage} [htbp]{5cm}
        \tabitem The buggy code fails to throw an IllegalArgumentException if the values of the method at the three points have the same sign. \\
        \tabitem The developer patch fixes the bug by inserting conditional statements to checking the values.
    \end{minipage}
    \\
    \hline

    Chart-7 &
    \ttfamily
    \setlength\tabcolsep{2pt}
    \begin{tabular}{lll}
        1 & \textcolor{red}{-} & long s = getDataItem(this.minMiddleIndex).getPeriod().getStart() \\
        2 & \textcolor{green}{+} & long s = getDataItem(this.maxMiddleIndex).getPeriod().getStart() \\
        3 & & .getTime();\\
        4 & \textcolor{red}{-} & long e = getDataItem(this.minMiddleIndex).getPeriod().getEnd() \\
        5 & \textcolor{green}{+} & long e = getDataItem(this.maxMiddleIndex).getPeriod().getEnd() \\
        6 & & .getTime(); \\
    \end{tabular}
    &
    \begin{minipage} [htbp]{5cm}
        \tabitem The buggy code returns the wrong index of the start (end) time and throws an AssertionFailedError. \\
        \tabitem The developer patch fixes the bug by changing related parameters in called methods.
    \end{minipage}
    \\
    \hline

    Chart-12 &
    \ttfamily
    \setlength\tabcolsep{2pt}
    \begin{tabular}{lll}
        1 & & super(); \\
        2 & \textcolor{red}{-} & this.dataset = dataset; \\
        3 & \textcolor{green}{+} & setDataset(dataset); \\
    \end{tabular}
    &
    \begin{minipage} [htbp]{5cm}
        \tabitem The dataset is not wired to a listener,  when dataset is passed into constructor for MultiplePiePlot. \\
        \tabitem The developer patch fixes the bug by setting the new dataset, and registering the chart as a change listener.
    \end{minipage}
    \\
    \hline
  \end{tabular}
\end{table*}

\section{Research Questions}

In this study, we aim to investigate the following three research questions.

\begin{description}
    \item[RQ1.] How do state-of-the-art APR tools perform comparing with manual program repair?
        \\
        To answer RQ1, we compare the number and quality of patches generated by manual program repair and several state-of-the-art APR tools.
        We also distinguish how the characteristics and type of bugs impact the two repair scenarios.
        Further, we analyze the repairing efficiency in terms of repair time and the number of candidate patches.

    \item[RQ2.] Are state-of-the-art APR tools beneficial to manual program repair as debugging aids?
    	\\
        To answer RQ2, we discuss whether the assistance of APR tools (i.e., repair reports) can further boost the performance of manual program repair.
        Specifically, we also investigate how manual program repair is influenced by different factors (e.g., patch generation techniques, report formats)
    \item[RQ3.] What are participants' opinions on the use of state-of-the-art APR tools?
    	\\
        To answer RQ3, we perform a questionnaire  survey to qualitatively analyze participants' opinions on the assistance of APR tools, such as the positive and negative thoughts.
  \end{description}

\section{Empirical Study}

In this section, we enumerate the subject programs, participants and state our study design in detail.

\subsection{APR tools}
\label{apr_selection}
Although we initially plan to consider all repair tools proposed in the last decade \cite{2020Benton}, we are limited by the fact that the size of selected tools may lead to endless human study.
Thus, following prior work \cite{2014Tao,2019Durieux,2020Chen}, we systematically consider several representative tools in our experiment based on the following criteria. 

\begin{enumerate}
	\item All selected tools are required to be publicly available and possible to run, as we need to run the tools in the experiment.
	\item The number of selected tools should be proper,
	  as our human study involving too many tools may be unaffordable (e.g., two tools are used in \cite{2014Tao}). 
	\item The selected tools can represent all possible categories, as we want to consider all repair techniques in the experiment. 
	\item The tool should require only the source code of buggy program and its corresponding test suite, as the two elements are the two inputs specified in the problem statement of APR \cite{2019Durieux}. Thus, according to recent empirical studies \cite{2020Liu,2020Benton}, all learning-based tools are excluded (e.g., \cite{2021Jiang, 2020Lutellier}).
\end{enumerate}
 





Accordingly, we select three repair tools, each representing the state-of-the-art technique in the corresponding category.
Specifically, the selection configuration is the same as a recent APR work \cite{2020Chen} except PraPR, because only source-level APR tools are considered in that work.
The first one is a template-based tool, \textbf{PraPR}, which is proposed recently and fixes bugs at the bytecode level.
Compared against other state-of-the-art tools, PraPR is able to fix more bugs with a much lower overload and is considered as one of the most advanced repair tools.
The rest two are a heuristic-based tool \textbf{SimFix} and a constraint-based tool \textbf{ACS}.
Both of them are proposed in the past few years and able to fix the most number of bugs compared to all the studied tools for Defects4J dataset when published.
In brief, these selected tools not only represent all possible categories, but also cover other different aspects (e.g., whether one most plausible patch or all plausible patches are returned, and whether generated patches can be applied directly or there is a need to translate from an intermediate representation, etc.).
As such, this variance in different can help provide reliable results in our analysis.

\subsection{Subject Dataset and Bugs Selection}
There exist several benchmarks in recent APR literature \cite{2017Lin, 2018Saha, 2019Madeiral}.
After searching the literature for benchmarks, we adopt Defects4J \cite{2014Just}, as it has been continuously developed for a long time and has become the most widely studied dataset in APR studies \cite{2019Ghanbari, 2019Lou, 2020Liu, 2020Benton}, or even other software engineering research (e.g., fault localization \cite{2018Allamanis, 2016B, 2017Pearson, 2013Qi} and test case prioritization \cite{2018Miranda, 2019Cruciani}, etc.) in general.
Defects4J consists hundreds of known and reproducible real-world bugs from a collection of 16 real-world Java programs.

Note that we are unable to successfully apply ACS and SimFix to the new subjects (e.g., Mockito project) beyond the original paper \cite{2017Xiong, Jiang2018}.
SimFix is unable to locate reusable code snippets in new subjects and ACS is no longer allowed to use programmed queries in GitHub\footnote{https://github.com/.} due to the changed interface.
Thus, according to the recent APR study \cite{2020Benton}, the remaining four subjects (i.e., Chart, Time, Lang, and Math projects) are used in our experiment.

In recent work, Durieux et al. \cite{2019Durieux} show that APR tool may overfit Defects4J dataset in terms of repairability.
Thus, instead of adopting the entire dataset, we carefully select several representative bugs to mitigate the overfitting problem that satisfy the following criteria: 

\begin{enumerate}
	\item The number of total selected bugs should be proper, to control the scale of human study (e.g., 5 bugs are used in \cite{2014Tao}). 
	\item The diversity of bug types is preferred, to represent the whole dataset \cite{2018Sobreira}. 
	\item The performance distribution of APR tools on the whole selected bugs should be balanced, to avoid single APR tool may overfit the whole selected bugs. 
	\item The performance distribution of APR tools on each single selected bug should be varied, to avoid single bug type may overfit the whole selected tools.
\end{enumerate}

Accordingly, we randomly select two bugs for each subject, resulting in a total of eight real bugs.
The detailed information is summarized in Table \ref{Table1},
where column "$BugID$" presents the version of the buggy subject, and the remaining two columns present the developer patch and corresponding description of the bug.
Specifically, these selected bugs manifest different symptoms and cover various bug types.
There exist four bugs involving conditional statements (i.e., Time-15, Lang-10, Lang-7 and Math-73), two bugs involving method calls (i.e., Time-7, Chart-7), and two bugs involving assignments and return statements  (i.e., Math-5, Chart-12).
The distribution of bug types is consistent with that of  Defects4J \cite{2018Sobreira}, which indicates that they can represent the whole dataset well.
Meanwhile, we conform all APR tools perform similarly for the whole selected bugs (e.g., each tool can correctly, plausibly and abortively fix 1/3 bugs approximately).
Besides, APR tools perform diversely for each bug (e.g., each bug is correctly-fixed, plausibly-fixed and abortively-fixed by the three tools respectively) , to ensure each bug type can provide various report outcomes.




\subsection{Participants}
There are 20 participants involved in our experiment\footnote{It is worth noting that all participants are aware of the experiment and agree to the use of relevant data.}. 
All of them are carefully selected from graduate-level software engineering classes.
Besides, as a part of their course work, they are familiar with software testing and debugging.
It is noteworthy that we confirm the selected students' backgrounds represented a full range of experiences.
Specifically, some students are professional developers with rich experience in industry (and return to school), others have one or more internships in companies, and yet others have limited programming experiences outside of school.
As such, we are comfortable with this variance in experiences, which can help to provide reliable results in our analysis.



\subsection{Experiment platform}
We have developed Mooctest\footnote{http://mooctest.net/.} ,
an online platform for a Coursera course on software testing,
which can provide all supportive features and consists of three major modules:

\begin{enumerate}
	\item Arrangement module is able to arrange bugs and corresponding test suite for all participants automatically.
	\item Submission module is able to record all patches submitted by participants.
	\item Evaluation module is able to identify whether the submitted patch can pass the available test cases or not.
\end{enumerate}

Unlike the study design in \cite{2014Tao}, we schedule a repair task for one bug individually, as putting all bugs together may influence participants' debugging behaviour.
They may be free to skip any bugs, which is inconsistent with a practical debugging scenario \cite{2016Xie}.

\subsection{Evaluation metrics}
We measure the repair performance by the effectiveness (i.e., repairability and correctness) and efficiency (i.e., debugging time and the number of generated patch candidates).
 We also qualitatively investigate participants’ opinions on using APR tools as debugging aids by the survey feedback.

\textbf{Repairability} focuses on the number of correct patches.
A patch is considered to be correct if it can pass all test cases and be semantically equivalent or similar to the corresponding developer patch (described in Table \ref{Table1}).

Thus for the plausible patches (checked by our platform automatically), two volunteers with 7-year Java programming experience conduct three rounds of manual inspection to identify whether the patches are correct or not, according to the previous work \cite{2014Tao}.

\textbf{Correctness} indicates whether a generated patch is correct or not.
It is widely considered that patches generated by an APR tool should have a high probability to be correct, 
while the overfitting problem (i.e., the generated patch is plausible but overfitting) is still a major and long-standing challenge \cite{2019Gazzola}.



\textbf{Debugging time} is recorded automatically.
For manual program repair, debugging time repair is calculated by the time elapsed from each repair task beginning until the patch submission.
For automated program repair, debugging time is calculated by the repair report (e.g., Fig. \ref{fig:report}).

\textbf{The number of patch candidates} indicates how many attempts automated (manual) program repair makes before fixing the bug.
This metric is independent of a few redundant variables (i.e., machine configurations) that are unrelated to the approach implemented in APR \cite{2020Liu}, and can provide reliable results to supplement efficiency evaluation in our study.

\textbf{Feedback} is collected from the questionnaire survey.
The survey is conducted online and all participants can submit their free-form answers on condition of anonymity to mitigate some risks, which could bias the results (e.g., participants might feel that the answers could affect their performance).

\subsection{Experimental procedure}
\begin{figure}[t]
\centering

\graphicspath{{graphs/}}
    \includegraphics[width=0.47\textwidth]{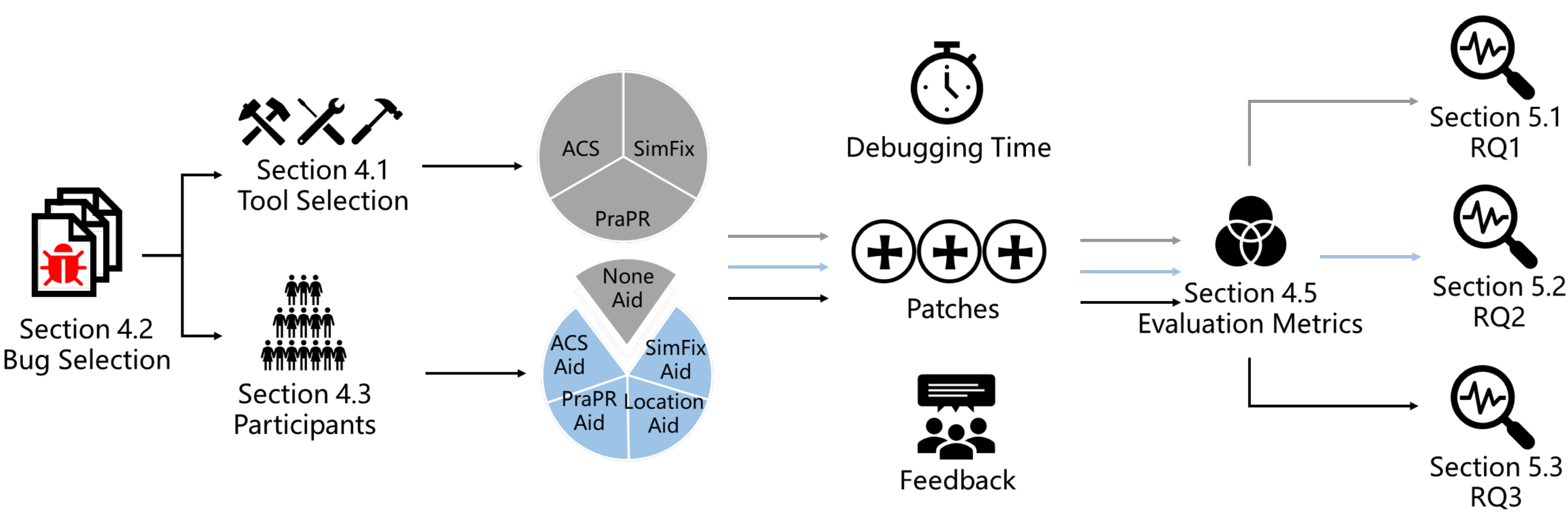}
    \caption{Overview of the experimental process}
    \label{fig:process}
\end{figure}

Firstly, an introduction to APR tools and our experimental platform in the form of document and presentation video is given to all participants, to ensure they fully understand the experimental procedure.
Then, to mitigate the effect of our platform's usability, we conduct a pre-test involving an additional real bug from Defects4J.
Finally, participants are required to fix all eight real bugs independently within four hours.
Fig. \ref{fig:process} presents an overview of the experimental procedure.
As a reminder, the restriction on repair hours is imposed to avoid endless debugging.
Based on participants' performance in the pre-test and study designs in previous studies\cite{2016Xie, 2014Tao},
4 hours should be adequate for participants to complete repair tasks
at a reasonably comfortable pace.

Specifically, to answer RQ1, we compare the repair performance of the participants with no aid against three state-of-the-art APR tools.
To answer RQ2, we compare the repair performance of the participants with no aid against the participants with APR aids (i.e., repair reports produced by the tools).
However, as repair reports already reveal suspicious code elements, for a fair comparison, we provide buggy location reports to the control group instead of leaving it completely unaided.
The buggy location report is generated by an off-the-shelf fault localization technique Ochiai \cite{2007Abreu}, which is widely adopted in recent APR studies (e.g., PraPR, SimFix and ACS) \cite{2019Ghanbari, 2018Xiong, Jiang2018}.
For simplicity, participants are rated in four levels based on programming experience, and then participants within the same level are randomly divided into five groups: None group (unaided), PraPR group (PraPR aided), ACS group (ACS aided), SimFix group (SimFix aided), and Location group (buggy location aided), resulting in a total of 160 manual program repair tasks (i.e., 4 participants and 8 bugs for each group).
Thus, the first group (32 repair tasks) is adopted to answer RQ1 and the other groups are adopted to answer RQ2.
Besides, to answer RQ3, all the participants attend an online survey to submit their opinions on the assistance of APR tools.

\section{Results and analysis}
In this section, we will analyze our experimental results to address the three research questions.

\subsection{RQ1: Comparison between automated program repair and manual program repair}
\label{rq1}
To answer RQ1, we measure the repair performance of three state-of-the-art APR tools and manual program repair (participants from None group) in terms of effectiveness and efficiency.
Thus, there exist 32 repair tasks performed by participants and 24 repair reports generated by APR tools for the eight bugs.
It is noteworthy that although some criteria (e.g., repairability \cite{2019Durieux} and efficiency \cite{2020Liu}) have been studied in the literature, there still exist little systematic work to compare APR tools with manual program repair.

\begin{table}[t]
\centering
 \caption{The effectiveness comparison between automated and manual program repair}
  \label{manual_automated}
    \begin{tabular}{|c||c|c|c|c|}
     \hline
        Program & Participants & PraPR & SimFix & ACS	\\
     \hline \hline
        Time-7 & 0(0) & 0(0) & 1(1) & 0(0)	\\
        Time-15 & 2(2) & 0(0) & 0(0) & 1(1)	\\
        Lang-10 & 0(0) & 1(1) & 0(1) & 0(0)	\\
        Lang-7 & 4(4) & 0(5) & 0(0) & 1(1)	\\
        Math-73 & 2(3) & 0(4) & 0(1) & 0(1)	\\
        Math-5 & 4(4) & 1(3) & 1(1) & 1(1)	\\
        Chart-7 & 4(4) & 0(16) & 1(1) & 0(0)	\\
        Chart-12 & 1(1) & 1(2) & 0(1) & 0(0)	\\
        All & 17(18) & 3(27) & 3(6) & 3(4)	\\
     \hline
  \end{tabular}
\end{table}


\subsubsection{Effectiveness}

We first analyze the repairability (i.e., the number of correctly-fixed bugs) of automated and manual program repair overall.
Then, for each bug, we distinguish how the types of bugs impact automated and manual program repair.
We also investigate whether there exists a difference in patch correctness and code quality between automated and manual program repair.

Table \ref{manual_automated} presents the repair information,  where per row represents all four repair scenarios for each bug (i.e., PraPR, SimFix, ACS and participants).
Specifically, each cell is represented as $x(y)$, where $x$ is the number of correct patches and $y$ is the number of produced plausible patches by participants (APR tools).
It is noteworthy that the submissions will not be counted if no changes are made by participants.

\textbf{Repairability Evaluation.}
Overall, participants can fix six of all eight bugs, twice as that of each APR tool.
However, a combination of the three tools can fix more bugs (except for Math-73) than that of participants.
We further analyze how the bug types impact the repair performance of automated and manual program repair.

 As repair action regarding condition is able to fix half of the bugs in existing approaches \cite{2017Xiong}, we firstly focus on the bugs (i.e., Time-15, Lang-10, Lang-7 and Math-73) involving conditional statement.
Specifically, for the bugs involving conditional statement (i.e., Time-15, Lang-10, Lang-7 and Math-73),
participants can fix three of them, except Lang-10.
In fact, participants need to remove the corresponding conditional statements without adding any statements when attempt to fix bug Lang-10.
It is rare to perform only removal actions in repair tasks, as most bugs are fixed by modification or addition actions.
We confirm that there are fewer than 1\% (2/395) bugs requiring only removal actions in Defects4J \cite{2018Sobreira}.
Thus, participants may ignore such low-frequency repair actions due to their unconscious mind.
Similar to manual program repair, ACS cannot correctly fix Lang-10, as it aims at generating precise conditions.
Meanwhile, SimFix fails to generate any correct patches for all these bugs, as no similar code snippets containing correct repair actions are found.
For example, though there is one code snippet similar to the inserted statement for bug Time-15, their context varies greatly.
Thus SimFix cannot extract such code snippets, which may be the reason why this tool is unsatisfactory.
Besides, PraPR can only fix Lang-10 by RemoveConditionalMutator and cannot generate any additional conditional statement with correct predicates for the other three bugs.

It is also observed that participants' performance varies in the bugs involving method call (i.e., Time-7, Chart-7).
Intuitively, the edit distance for Chart-7 is 24, and the one for Time-7 is 132.
An explanation may be that fixing Time-7 needs to modify the called method and corresponding parameters, where it is difficult to search the right parameters.
While the repair action on Chart-7 is simple, all participants are able to fix the bug.
It is also rough for mutation-based tool PraPR to fix Time-7, which needs two chunks (i.e., a sequence of continuous changes).
Participants and PraPR can fix all the remaining bugs involving assignments and return statements (i.e., Math-5, Chart-12).
However, ACS and SimFix cannot fix Chart-12 because of the primary design mechanism.

Different APR tools are suitable for the bugs with corresponding types because of their design mechanisms.
In contrast, participants can always repair all types of the bugs with a deeper understanding of the program, although they may ignore some low-frequency repair actions (e.g., removal actions in Lang-10) due to the unconscious mind sometimes.

\textbf{Correctness Evaluation.}
As prior work \cite{2019Gazzola,2010Gu,2018Monperrus,2015Qi,2018Yi} confirms that APR tools may generate tting patches (i.e., patches passing  the  entire  available test  suite  may  not  generalize  to  other  potential  test  cases),
we thus investigate whether there are differences in the patch correctness between automated and manual program repair.

As shown in Table \ref{manual_automated}, the correctness of manual patches is over 90\% (17/18), which is far higher than that of auto-generated patches (about 25\% $\approx$ 9/37).
APR tools usually use the available test suite to verify the generated patches.
However, a patch passing the available test suite may not generalize to other potential test cases.
APR tools may produce many incorrect patches with insufficient test cases, even for ACS, which aims to achieve high precision.
Instead, participants can identify the generated patch correctness by human ability in comprehending the source code (e.g., identify the functionality of the buggy method from Javadoc comments).

The correctness of auto-generated patches heavily depends on the quality of the available test suite, while participants 
generate correct patches with a deeper understanding of the source code.

\begin{figure}[t]
\centering
\ttfamily
    \begin{tabular}{p{0.5cm}p{0.08cm}p{7cm}}

    \hline
    \multicolumn{3}{l}{\textbf{Patch $P_1$ Generated by the participant}} \\
        1 & & if (str.startsWith("--")) \{ \\
        2 & \textcolor{red}{-} & \qquad return null; \\
        2 & \textcolor{green}{+} & \qquad throw new NumberFormatException(str + "is not a valid number"); \\
        3 & & \} \\

    \hline
    \multicolumn{3}{l}{\textbf{Patch $P_2$ Generated by
    ACS}} \\
        1 & & if (str.startsWith("--")) \{ \\
        2 &\textcolor{green}{+} & \qquad if (str.startsWith("--")==true) \{ throw new NumberFormatException();\}\\
        3 & & \qquad return null; \\
        4 & & \} \\
    \hline
    \end{tabular}
\caption{Generated patches for Lang-7}
\label{patch_quality}
\end{figure}
\textbf{Quality Evaluation.}
As prior work suggests that generated patches are less likely to be accepted by developers due to poor readability and maintainability \cite{2013Kim,2012Fry},
we thus investigate whether there are differences in the patch quality between automated and manual program repair.

For patches generated by automated and manual program repair, the former may be less readable, while the latter is more like the developer patch.
For example, Fig. \ref{patch_quality} illustrates the correct patches for bug Lang-7 generated by the participant and ACS.
Both of them are different from the developer patch.
However, the two patches are semantically equivalent and can fix the bug correctly.
As is shown in Table \ref{Table1}, the developer patch calls the method \textit{trim()} in the method \textit{createBigDecimal()},
while the method \textit{createBigDecimal()} will only be called by buggy method \textit{createNumber()}, which ensures that the parameter "\textit{str}" does not contain any space.
As a result, calling \textit{trim()} is not necessary and the two patches in Fig. \ref{patch_quality} are identical with developer patch in Table \ref{Table1}.
In fact, patch $P_1$ modifies the return statement and inserts an exception.
In contrast, patch $P_2$ inserts a new conditional statement semantically equivalent to line 1.
Thus, the entered string "\textit{==true}"  and original return statement "\textit{return null;}" are almost meaningless, and even misleading for human.

Patches generated by manual program repair are more readable than those by APR tools, and there may exist redundant code in the latter.
As a result, although APR tools can generate some correct patches, they still are required reexamination for deployment into the application due to its quality.
This calls for future research on automated patch transformation to normalize automatically generated patches.

\subsubsection{Efficiency}

\begin{table}[t]
    \centering
    \caption{The efficiency comparison between  automated and manual program repair}
    \label{RQ1-time}
    \begin{tabular}{|c||ccc|ccc|}
        \hline
        \multirow{2}{*}{Type} & 	\multicolumn{3}{c}{\#Patch} 	&
        &	Time(s)	&	\\
       \cline{2-7}
        &	min	&	avg	&	max	&	min	&	avg	&	max		\\
        \hline \hline
        Participants	&	\textbf{1}	&	\textbf{1.70} 	&	6	&	\textbf{315} 	&	\textbf{1072.00} 	&	\textbf{1800} 	\\
        PraPR	&	207	&	960.30 	&	1523	&	29 	&	54.00 	&	79 	\\
        SimFix	&	\textbf{1}	&	544.00 	&	2054	&	60 	&	656.00 	&	1600 	\\
        ACS	&	\textbf{1}	&	2.00 	&	\textbf{5}	&	76 	&	149.80 	&	288 	\\
        \hline
    \end{tabular}
\end{table}

We compute debugging time to analyze the efficiency of APR tools against manual program repair.
However, debugging time criterion may not be a sufficient metric for efficiency in our experiment due to the inherent difference between human and machine.
Thus, according to existing work \cite{2020Liu}, we also adopt the number of generated patch candidates to supplement efficiency evaluation in our study.

Table \ref{RQ1-time} provides statistical information about debugging time and the number of patch candidates.
In this table, column "\textit{\#Patch}" presents the \textit{minimum/average/maximum} number of all patch candidates, as well as the time cost for column "\textit{Time(s)}".
It is observed that manual program repair always consumes the more debugging time (1985\%, 163\% and 719\%) than APR tools on average.
On the contrary, it seems that manual program repair needs fewer average attempts (1.7) to generate a correct patch than those of all APR tools (960.3, 540.0 and 2.0).
Specifically, participants usually need to understand the functionality of the buggy method and before attempting to fix it.
In contrast, APR tools can use heuristic-based methods to generate a mass of program variants based on various variation rules, and then adopt a test-driven approach to find plausible patches.
As a result, manual program repair may require a significant amount of time in the comprehension process, and APR requires lots of attempts in the verification process.
Although such conclusions may be unsurprising and confirmed intuitively, such a confirmation warrants rigorous and substantial experiments.

Participants usually require more debugging time to fix a bug owing to the comprehension process, while APR may need more attempts to generate a plausible patch.
Another interesting finding is that manual repair cost more than all 3 tools combined, while fixing less bugs than the 3 tools.

\finding{1}{
Overall, our comparison between automated and manual program repair reveals that
(1) manual program repair habitually ignores the repair actions with a low frequency;
(2) the repair tool is effective for the bugs with certain types due to their design mechanisms;
(3) patches generated by manual program repair has a higher correctness ratio and quality than APR;
(4) APR tools usually require less debugging time, while they need more attempts to generate a plausible patch.
}

\subsection{RQ2: The assistance of state-of-the-art APR tools}

Specifically, we first analyze repairability of participants from four groups (i.e., Location, SimFix, ACS and PraPR group) to judge whether APR aids (i.e., repair reports) can improve participants' performance.
Then we analyze how the factors (e.g., the report type and patch quality) impact participants' performance, so as to further contribute to improving the state-of-the-arts.


\begin{figure}[!t]
\centering
\graphicspath{{graphs/}}
    \includegraphics[width=0.48\textwidth]{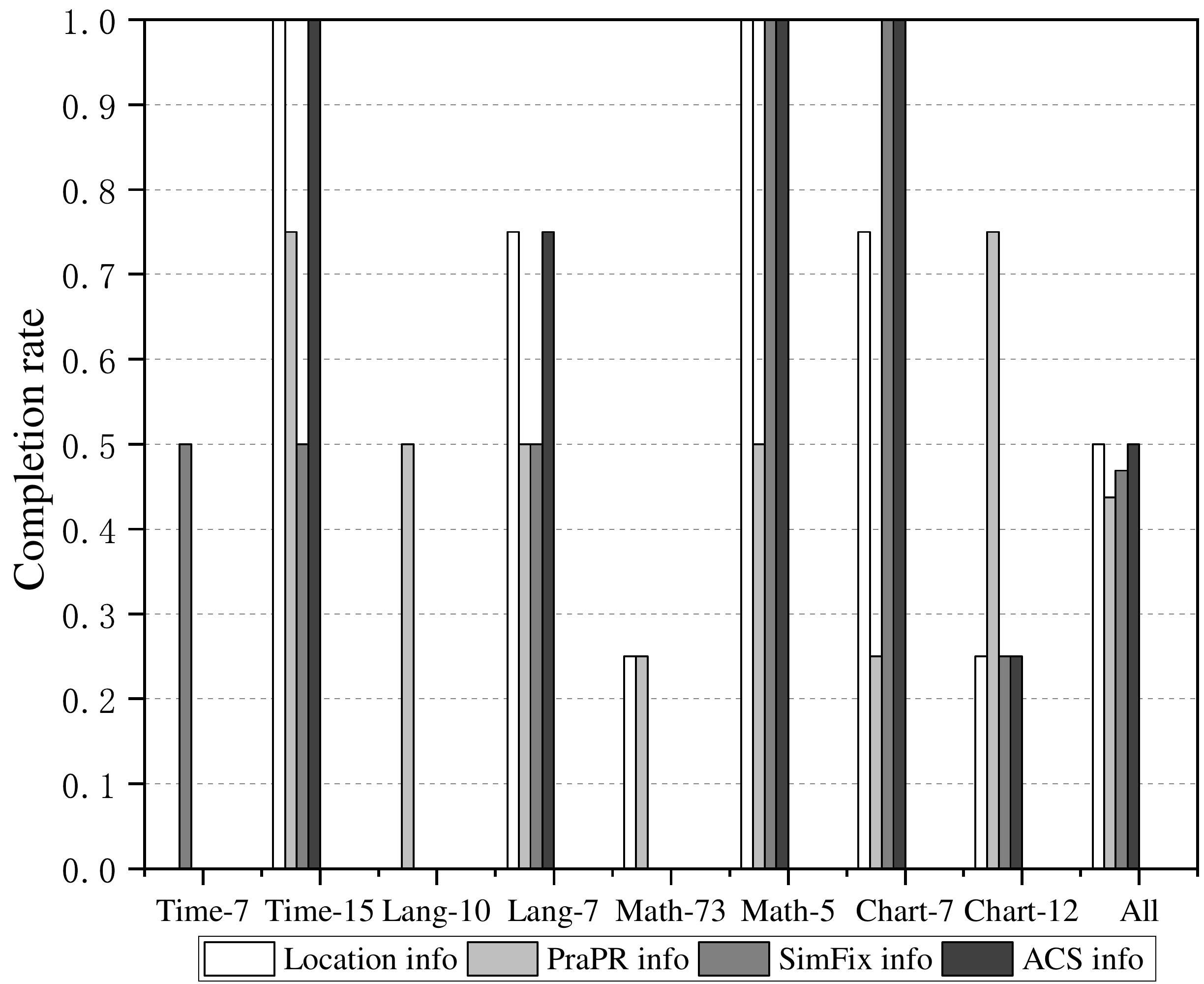}
    \caption{Completion ratio for all bugs}
    \label{RQ2-1}
\end{figure}

\subsubsection{Overall performance}

For simplicity, to evaluate repairability of four groups, we use completion ratio $C_r$,  defined as $P_n / T_n$, where $P_n$ is the number of participants who have correctly fixed the bug within the stipulated time (i.e., the number of correct patches), and $T_n$ is the total number of all participants taking part in this group (i.e., 4 participants).

Fig. \ref{RQ2-1} demonstrates the comparison for all eight bugs, where the four bars per bug represent four groups, resulting in a total of 32 repair types (8 bugs * 4 groups).
There are a total of nine repair types achieving the highest $C_r$ in each bug, eight of them from APR groups, and one of them from Location group.
Meanwhile, Time-7 and Lang-10 can be fixed only if SimFix and PraPR reports are provided.
It seems that the group with repair reports has a high chance of achieving higher $C_r$ and participants may fix some previously unrepaired bugs with appropriate repair reports.
However, there are also a total of 17 repair types achieving the lowest $C_r$, 10 of them from APR groups, and only 4 of them with Location group.
Meanwhile, participants can fix Math-73 with the buggy location report, while no repair tasks are successful with SimFix and ACS reports.
It seems that some repair reports may also decrease $C_r$ and even mislead participants resulting in failed repair tasks.


In other words, manual program repair overall can benefit from state-of-the-art APR tools in terms of the number of correctly-fixed bugs, while in some cases, such tools may also have a significant adverse effect.

\subsubsection{Performance impacted by the tool types}
Existing APR tools adopt various patch generation techniques based on different mechanisms.
For example, ACS focuses on condition synthesis and SimFix relies on similar code snippets.
Also, both of them return only one most plausible patch in source-code level, while PraPR returns all plausible patches in bytecode level.
Based on these insights, we further investigate the manual program repair performance impacted by APR tools.

\begin{figure*}
\centering
\subfigure[The patch correctness for three repair tools] {
 \label{tool-correctness}
\includegraphics[width=0.45\columnwidth]{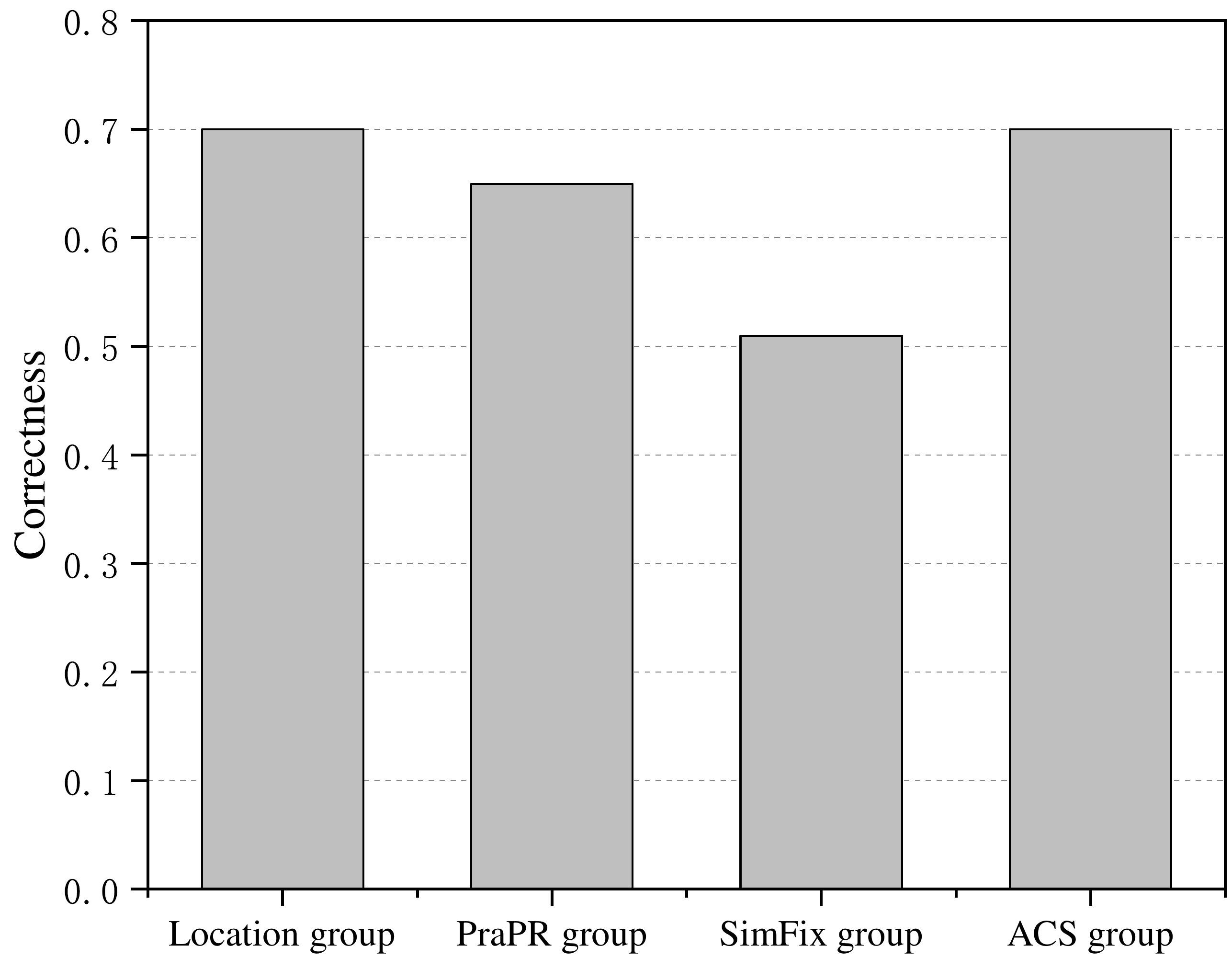}
}
\subfigure[The debugging time for three repair tools] {
\label{tool-time}
\includegraphics[width=0.45\columnwidth]{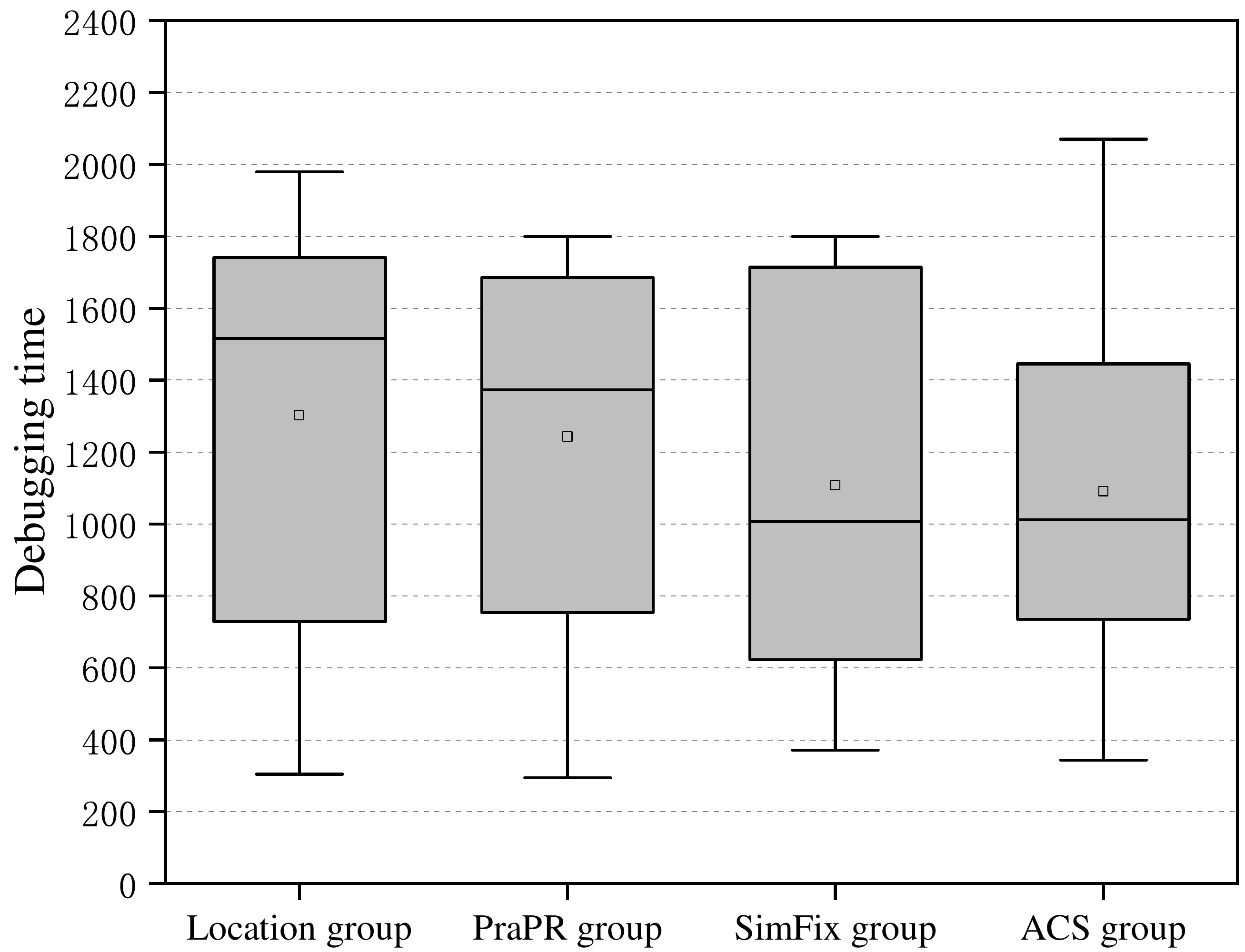}
}
\subfigure[The patch correctness for three report types] {
\label{info-correctness}
\includegraphics[width=0.45\columnwidth]{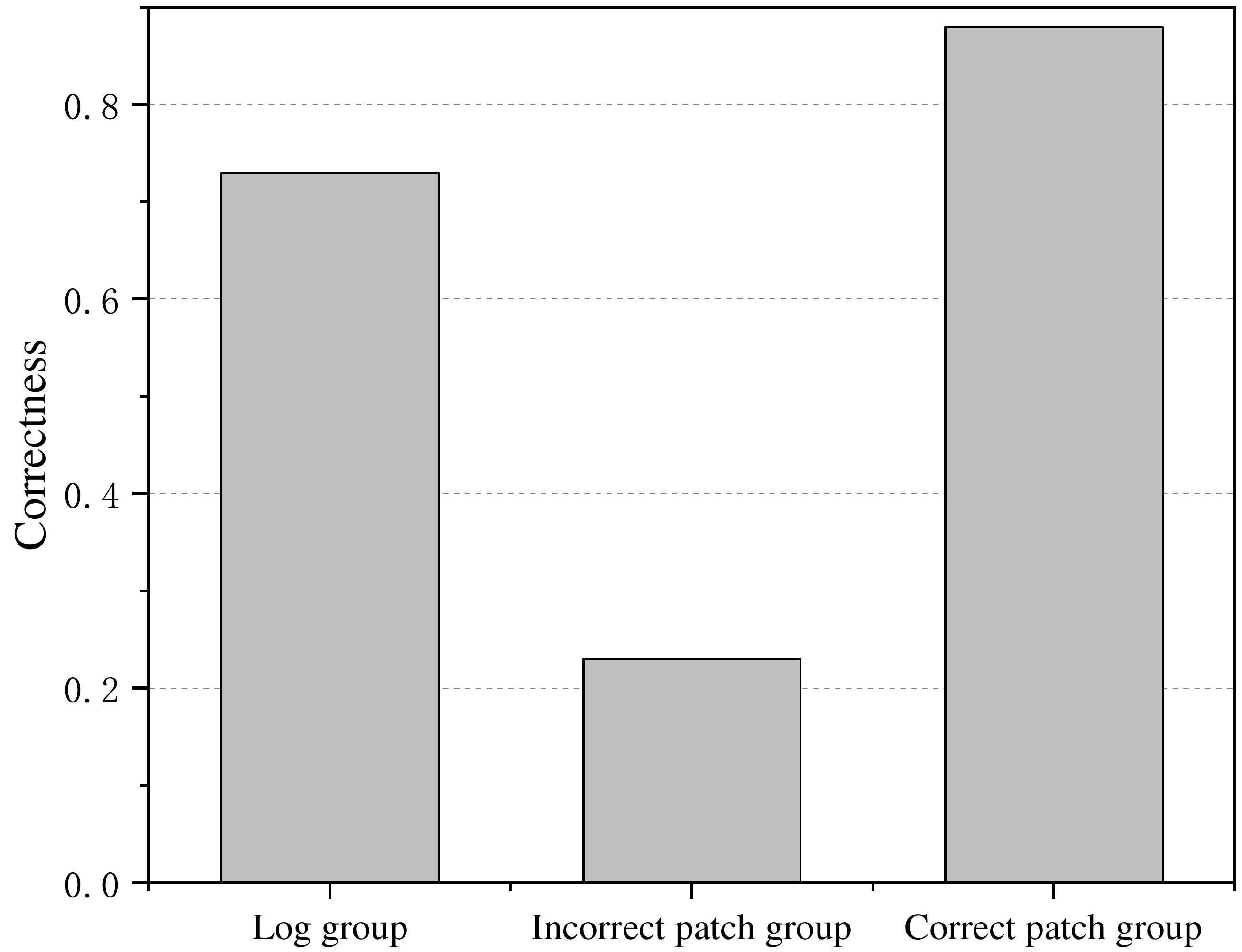}
}
\subfigure[The debugging time for three report types] {
\label{info-time}
\includegraphics[width=0.45\columnwidth]{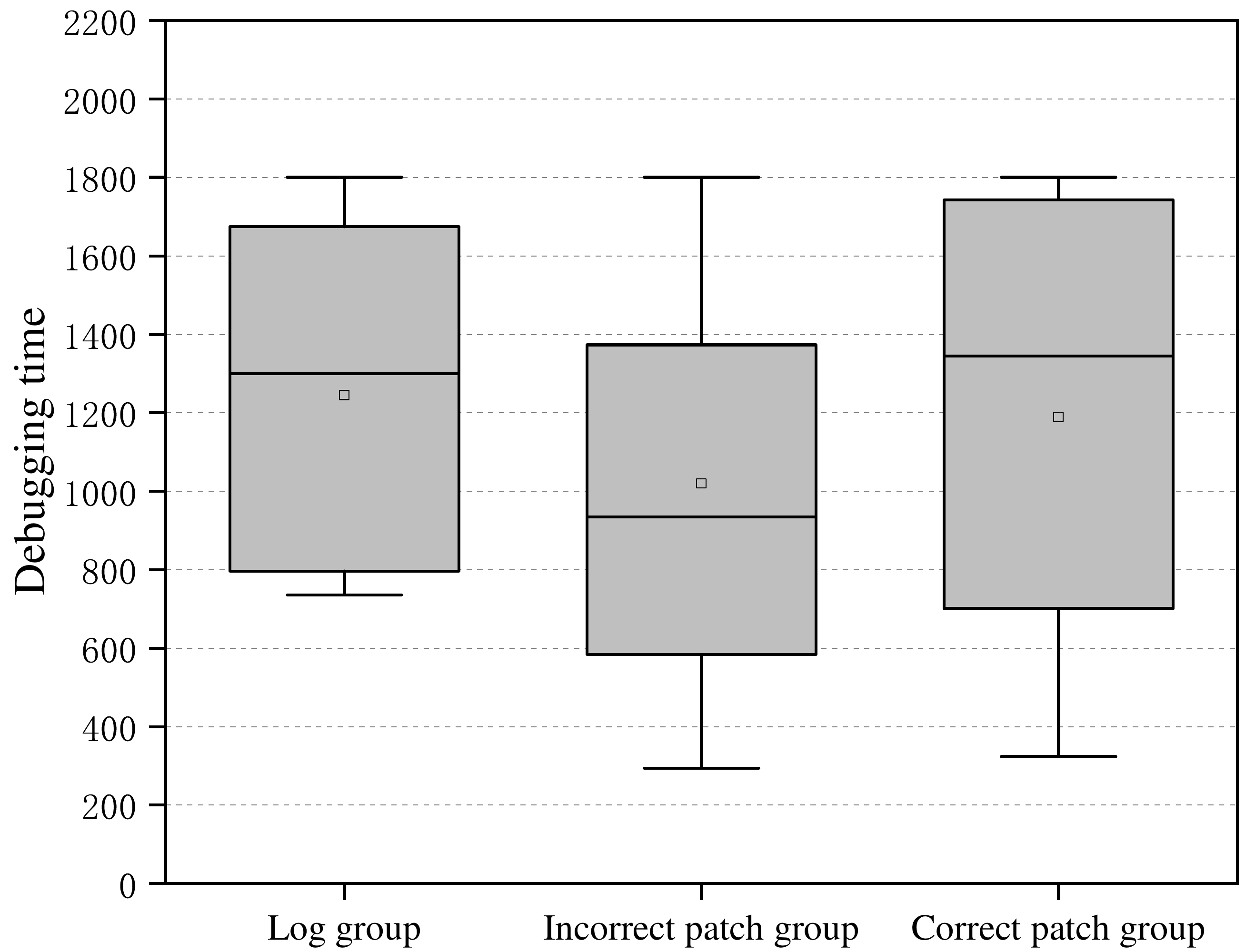}
}
\caption{ The performance for different repair tools and reports }
\label{RQ2-tool}
\end{figure*}

Fig. \ref{tool-correctness} shows the percentage of correct patches for the groups.
Specifically, patches submitted by the Location group are 70\% (16/23) correct, the same as ACS group.
The percentage of correct patches decrease to 51\% (15/29) for SimFix group.
PraPR reports do not improve correctness too, with only 65\% (15/23) patches being correct.

Fig. \ref{tool-time} shows the debugging time for the groups.
Location group has an average debugging time of 21.8 minutes, which is slightly slower than 20.8, 17.8, and 18.8 minutes for PraPR, SimFix and ACS groups, respectively.

It seems that repair reports may reduce participants’ debugging time slightly, while patch correctness drops to an even lower point than that of the control group.
Meanwhile, the three repair tools have little influence on the patch correctness and debugging time.
%
%

\subsubsection{Performance impacted by the report types}
\label{section2:info_type}

The recent APR studies \cite{2019Lou, 2019Ghanbari} demonstrate that, even state-of-the-art APR tools can only fix a small ratio of real bugs (i.e.,$<$20\% for Defects4J) fully automatically and are simply aborted for the vast majority of unfixed bugs.
Based on the repair outcomes of APR tools, the repair reports may contain different types of information (i.e., log and patch output), which could impact participants' debugging strategies and performance.
For example, the repair report only contains log output for the vast majority of unfixed bugs.
Thus, we want to investigate whether APR tools can also provide useful debugging information to help manual program repair even for the bugs that are hard to fix automatically.


Specifically, we divide repair reports into three types based on repair outcomes:
(1) log group: the corresponding repair task is only provided with a log output, including buggy statements and patches already tried.
(2) incorrect patch group: the corresponding repair task is also provided with a patch output, containing some plausible patches, which are not semantically equivalent to developer patch.
(3) correct patch group: the corresponding repair task is also provided with a patch output containing a correct patch, which is semantically equivalent to the developer patch.

As shown in Fig. \ref{info-correctness}, the log group has 73.3\% (11/15) patches being correct, which is similar to the control group.
The percentage of correct patches dramatically decreases to 23\% for the incorrect patch group.
On the contrary, the correct patch group performs much better than the above two groups, with about 88\% (28/31) patches being correct.
Meanwhile, the average debugging time for log group, incorrect patch group and correct patch group is 21.3, 17.0 and 19.8 minutes.

It is observed that the report types have a significant influence on the patch correctness, while debugging time is not affected.

\subsubsection{Performance impacted by the patch quality}
\label{section2:patch_type}
\label{section2:patch}
    \begin{table}[t]
        \centering
        \caption{The identification   results for all patches}
        \label{RQ3-patch}
        \begin{tabular}{|c||c|c|c|c|}
            \hline
            &	PraPR	&	Simfix	&	ACS	&	Sum	\\
            \hline \hline
            Correct	&	0.60/8.6	&	1.00/5.9	&	1.00/7.7	&	0.91/7.3	\\
            Incorrect	&	0.30/10.5	&	0.13/10.5	&	0.33/9.3	&	0.24/10.3	\\
            Sum	&	0.50/9.8	&	0.56/8.4	&	0.80/7.6	&	0.59/8.9	\\
            \hline
        \end{tabular}
    \end{table}
As discussed above, the repair performance increases when participants are aided by correct patches and decreases when participants are aided by incorrect patches.
Based on these insights, we want to investigate whether participants can accurately identify the patch correctness and how long it takes to complete the identification step.
Hence we require participants to answer questions about the given patch correctness (if provided in the repair task) and the identification time when they attempt to fix the bug.

Table \ref{RQ3-patch} summarizes the identification results.
Intuitively, each cell in the table is represented as $x/y$, where $x$ is the identification correctness and $y$ is the identification time.
As shown in Table \ref{RQ3-patch}, all participants can identify 59\% (26/44) patches correctly using an average of 8.9 minutes per patch.
The precision drastically increases to 91\% (21/23) when they are provided with correct patches and decreases to 24\% (5/21) when they are provided with incorrect patches.
The identification time (7.3 minutes) for correct patches is 41\% faster than incorrect patches (10.3 minutes).

Besides, among the three tools, patches generated by PraPR (9.8 minutes) consume more identification time than the ones generated by SimFix (8.4 minutes) and ACS (7.6 minutes).
One possible reason is that PraPR dumps each plausible fix to a class file and the users need to take advantage of a Java bytecode decompiler to decompile the resulting class file,
which might be time-consuming.
Meanwhile, the precision for all tools with correct patches is 100\%, except PraPR with 60\%.
Unlike the other tools, PraPR will provide all plausible patches to users (an average of 5 patches in our experiment), and it will be a little bit confusing to identify the patches as the amount increases.

Participants usually spend more debugging time on incorrect patches, while still easily misguided.
Thus the accuracy of auto-generated patches is particularly important in practice.
This observation urges a strict quality control for auto-generated patches if APR tools are used as debugging aids.

\finding{2}{
Overall, our analysis on the assistance of APR tools reveals that
(1) manual program repair can benefit from the assistance of APR tools in terms of the number of correctly-fixed bugs;
(2) the tool types have little impact on the correctness and debugging time;
(3) the report types have a significant influence on correctness and debugging time is not affected;
(4) plausible but incorrect patches have adverse  effects on the identification precision and identification time.
}

\subsection{RQ3: What are participants' opinions on APR tools}
\label{section:rq3}

 \begin{table*}[htbp]
 \centering
  \caption{Participants' positive and negative opinions on APR tools}
   \label{Tab:feedback}
     \begin{tabular}{|l|l|}
     \hline
     \textbf{Positive} & \textbf{Negative} \\
     \hline
     \hline
     \begin{minipage} [htbp]{0.45\textwidth}
     \emph{\textbf{It is effective to accelerate debugging process initially.}}\\
         \textbf{P1:} It helps to identify suspicious code elements quickly when developers are not familiar with the source code. \\
         \textbf{P2:} Participants could immediately recognize the buggy code elements and acquire the plausible solution to the problem without understanding all of the source code.\\
         \textbf{P3:} It can quickly identify buggy elements and further provide candidate fixes for developers to choose. \\
         \textbf{P4:} It is able to help me repair bugs faster.\\
     \end{minipage}
     &
     \begin{minipage} [htbp]{0.45\textwidth}
     \emph{\textbf{It is difficult to get started at the beginning.}} \\
         \textbf{N1:} It is difficult to adopt the tool at the beginning when developers are not familiar with it. \\
         \textbf{N2:} It is costly for developers to extract the content of the repair report. \\
         \textbf{N3:} An important point of the tool is to make the repair report easy for developers to understand. \\
         \textbf{N4:} The provided patch is generated based on frequent trial and error, without understanding the functionality of the code. \\
     \end{minipage}
     \\
     \hline
     \begin{minipage} [htbp]{0.45\textwidth}
     \emph{\textbf{It can provide multiple suspicious buggy code elements.}} \\
         \textbf{P5:} It can guide me to indentify potentially risky code elements. \\
         \textbf{P6:} I could be more likely to identify the buggy code elements when the patches are provided, despite most of them being incorrect. \\
         \textbf{P7:} It provide multiple buggy code elements, and  they are beneficial for me to repair the bug. \\
         \textbf{P8:} It can help me to identify buggy code snippets. \\
         \textbf{P9:} It provides accurant buggy locations, which is convenient for me to understand the bug. \\
         \textbf{P10:} It can identify the location where a bug may appear. \\
     \end{minipage}
     &
     \begin{minipage} [htbp]{0.45\textwidth}
     \emph{\textbf{It provide reports with a low accuracy.}} \\ \\
         \textbf{N5:} The accuracy of the repair report is not high, and it also suffer from poor readability and usability. \\
         \textbf{N6:} When developers are provided with the repair report, the accuracy should be improved. \\
         \textbf{N7:} The accuracy of buggy locations and the understandability of reports are low. \\
         \textbf{N8:} The accuracy of the tool is too low, and it even attempts to generate patches on code elements, which are obviously correct. \\
     \end{minipage}
     \\

     \hline
     \begin{minipage} [htbp]{0.47\textwidth}
     \emph{\textbf{It can provide useful patches.}} \\
     \\
     	\textbf{P11:} The key is that the tool can generates a usable patch. \\
     	\textbf{P12:} It can always provide patches and suggest useful guidelines for repairing. \\
     	\textbf{P13:} It can provide useful suggestions about how to repair the bug, and sometimes it can even provided the correct patch. \\
     	\textbf{P14:} It can provide patches and buggy statements. \\
     	\textbf{P15:} It can indentify buggy statements and sometimes even provide plausible patches directly. \\
     	\textbf{P16:} It can provide me with plausible patches and suspicious buggy statements. \\
     \\
     \end{minipage}
     &
     \begin{minipage} [htbp]{0.47\textwidth}
     \emph{\textbf{Its report is less understandable.}} \\
 	\textbf{N9:} For the tools that generate bytecode level patches, I hope the patches can be presented to developers in source-level to aid readability. \\
 	\textbf{N10:} When generating repair reports, tools should eliminate irrelevant information as much as possible to facilitate quick understanding. \\
 	\textbf{N11:} Tools should provide decompiled source-level patches if they are able to fix the bug, otherwise the suspiciousness value for each code element should be presented. \\
 	\textbf{N12:} Displaying abundant process data is of little significance to developers. \\
 	\textbf{N13:} Some repair reports have complex content, in fact, providing the most critical information is enough. \\
 	\textbf{N14:} Repair reports in bytecode format need to be decompiled in advance, as they are inconvenient for me to understand. \\
     \end{minipage}
     \\
     \hline
   \end{tabular}
 \end{table*}


To qualitatively investigate participants' opinions on employing APR tools,
we conduct an online survey, where all participants are required to submit their free-form opinions on APR aids.
Given that there is a lack of such human analysis in the existing literature, using the survey feedback in a practical debugging scenario is also meaningful.
In total, we receive 102 textual answers for several well-designed questions from 85\% of the participants.

\textbf{Are APR tools useful for manual program repair?} In general, 94.12\% of participants think manual program repair can benefit from APR tools, while only 5.88\% do not.
However, less than half (about 47.06\%) of participants declare that they are willing to use such tools in practice.
This gap reveals an urgent problem that although these tools have been well designed regarding some criteria, their usability needs to be further improved for development.

\textbf{Which patch generation strategy should APR tools adopt?} As it is observed that the amount of provided patches may influence the identification performance, we also conduct a corresponding questionnaire.
Overall, 58.82\% of participants think it is suitable for tools to provide all plausible patches and the others prefer only one patch with the highest accuracy.
This reveals APR tools need to achieve a trade-off between the accuracy and quantity of patches.

\textbf{What information is preferred when APR tools cannot provide any plausible patch?} 
So far, despite the success of recent APR tools, even the most advanced tool can only generate plausible patches for a small ratio of real bugs.
In other words, manual program repair cannot be provided with any patches for the vast majority of bugs.
Thus, we conduct a corresponding questionnaire to analyze what information is preferred when no plausible patch is provided.
In the end, 70\% of responses indicate a need for buggy location and patch execution information.
Although such information is provided in log output, it may be contained in thousands of lines of documents.
Participants may consume lots of time to extract key information, which is not worth compared to manual program repair completely unaided.
This reveals using the feedback (i.e., the log output) of APR to  refine an easy-to-read report for all possible bugs might be promising, where participants can know what attempts APR tools have made.


\textbf{Participants’ positive and negative opinions on the assistance of APR tools.}
Participants freely elaborated their views on the advantages and disadvantages of APR tools.
We divide them into six reasons why participants are positive and negative bout APR tools as debugging aids after an open coding phase \cite{1988Landauer}.
The summarization of these reasons along with participants' original answers are presented at Table \ref{Tab:feedback}.

Specifically, participants acknowledge that APR tools are able to accelerate debugging process at the beginning, as the tools can identify suspicious code elements instantly without comprehending all source code.
However, the assistance of APR tools may be costly because it is difficult to use the tools and extract the key contents from complex reports.
Another shared concern is that such tools simply use trial and error to generate a patch without comprehending the source code, so as to generate some abnormal patches.

There is a popularly accepted opinion that such tools can provide some useful guidelines (i.e., suspicious code elements and candidate patches) on how to fix the bugs.
However, provided guidelines do not guarantee that the debugging is going down the right path, since the reports can be confusing and misleading.

Interestingly, we also observe that almost half of the participants think the repair reports are not user-friendly.
For example, it might be a little bit confusing for participants to read a given patch in bytecode-level and time-consuming to extract useful information from documents with too much redundant information.
Hence some participants hope that irrelevant information can be eliminated to facilitate the extraction process.

\finding{3}{
Overall, our questionnaire survey reveals that
(1) there exists a huge gap between the repair performance in the benchmark and the usability in practice;
(2) achieving a trade-off between the patch accuracy and patch quantity is important in repair reports;
(3) the feedback in log output should be valued more in the future;
(4) participants are positive about APR tools to provide useful guidelines about buggy locations and even fix solutions, while also less confident about the accuracy and format of repair reports.
}

%

\section{Threats to validity}
\label{Threats to validity}
%
%

The selection of participants might be biased.
Due to a monetary limitation, following existing work \cite{2011Parnin, 2016Xie, 2015Feng, 2016Feng}, we recruit students instead of professional developers from industry, which may introduce a bias in our conclusions.
To mitigate this threat, we select participants with varied experience (i.e., formal work experience, internship work experience and limited work experience).
Meanwhile, Salman et al. \cite{2015Salman} report that both students and professional developers have similar performance for a new software engineering task.
As such, we believe the selection strategy may not be a key point to our user study.


With respect to the representativeness of subjects, all of them are selected from Defects4J.
It has been shown that APR tools may t the dataset in terms of repairability \cite{2019Durieux}.
Although we focus on the comparison between automated and manual program repair, a bias may still be introduced.
Thus, instead of adopting the entire dataset, we carefully select several representative buggy programs to mitigate the tting problem.
Meanwhile, to mitigate the threat that the used bugs and APR tools may not be representative of all bugs and repair techniques, we randomly select eight real bugs with varied symptoms and bug types and three state-of-the-arts from all possible APR categories.
As we have seen, this is the largest relevant study in APR field ever (e.g, there exist six bugs and two APR tools in \cite{2014Tao}).


The final threat to validity is that participants might blindly reuse provided patches instead of really fixing bugs on their own, if plausible patches are provided in repair reports (i.e., the patch output).
To prevent such behaviours, we emphasize in advance that provided patches may not be correct, and participants should make their own judgement.
We also require all participants to judge the patch correctness (if provided) before submissions, and discover that there exists a similar debugging time cost between participants with or without provided patches (described in section \ref{section2:patch_type})
It indicates that participants are less likely to reuse provided patches unconditionally, which may otherwise take only seconds to complete.


\section{Practical Guidelines}
Based on the observations in our experiment, we have learned essential aspects to consider when using APR tools in a real debugging scenario.
Now we summarize our suggestions on improving the practicality of APR tools.


Repair report should be presented in a more user-friendly way.
As discussed in Section \ref{rq1}, it is necessary to recheck generated patches for deployment due to the quality issue.
Moreover, it would be misleading for participants to identify patch correctness if they are provided with all plausible patches.
Thus, it is better to normalize automatically generated patches and achieve a trade-off between the accuracy and quantity of patch according to the practical situation.

Misleading repair information should be reduced.
Misleading information (e.g., incorrect patches in Section \ref{section2:patch}) may delay the repair process as participants will spend a lot of time verifying such information.
Meanwhile, half of the participants are less confident about the accuracy ratio of the provided repair reports.
This can be done by designing better algorithms to provide accurate information.
However, the remarkable progress of APR tools requires a long and continuous effort.
In a similar area of fault localization, Le et al. \cite{2015Le} suggested predicting the accuracy of fault localization information before utilizing them, which can be adopted for APR.
Also, we can combine multiple reports from different APR tools to provide developers with more accurate repair information.

The feedback in the log output should be observed.
As discussed in Section \ref{section:rq3}, many participants consider it hard to extract significant information (e.g.,  buggy statements and fixes already tried) in the log output.
A possible explanation is that these tools focus on generating as many plausible patches as possible in the patch output, and do not take other useful debugging hints in the log output into account.
Thus, there exists lots of redundant information in the log output.
In fact, it is observed that the feedback in the log output can provide useful guidelines for debugging \cite{2019Lou}.
It seems a promising direction to unify plausible patches in the patch output and execution information in the log output.
Such a flexible debugging approach can always provide developers with refined debugging information for all possible bugs.


\section{Conclusion}
We conduct a large-scale human study to compare automated with manual program repair, and further investigate whether the assistance of APR tools can benefit manual program repair.
Our experiment involving eight real bugs and 160 repair tasks indicates that manual program repair may be influenced by the frequency of repair actions.
Besides, APR tools are able to improve manual program repair in terms of the number of correctly-fixed bugs, while the patch correctness may be adversely affected.
Furthermore, it is confirmed that the assistance of APR tools is promising, while the accuracy and format of repair reports need improving.
Based on these observations, some guidelines on improving the usability of existing APR tools (e.g.,  the misleading information in reports and the importance of feedback) are provided.


In the future, we will conduct a series of research from the suggested directions, 
to explore how to make APR tools consumable from the developer's perspective.
And further, it would be interesting to investigate how APR tools can be integrated into the regular developer workflow (e.g., IDE plugins).


\ifCLASSOPTIONcompsoc
  \section*{Acknowledgments}
\else
  \section*{Acknowledgment}
\fi
This work is supported partially by National Natural Science Foundation of China (61932012).

\ifCLASSOPTIONcaptionsoff
  \newpage
\fi

\bibliographystyle{IEEEtran}
\bibliography{reference}

\end{document}